\documentclass[twocolumn,aps,prc]{revtex4}
\usepackage{epsfig}
\usepackage{color}

\newcommand{\ket}{\rangle}
\newcommand{\bra}{\langle}

\def\be{\begin{equation}}
\def\ee{\end{equation}}
\def\bea{\begin{eqnarray}}
\def\eea{\end{eqnarray}}
\def\bse{\begin{subequations}}
\def\ese{\end{subequations}}

\begin{document}

\title{Sum rule study for Double Gamow-Teller states}

\author{H. Sagawa$^{1,2}$ and T. Uesaka$^{1}$} 
\affiliation{$^1$ RIKEN, Nishina Center for Accelerator-Based Science,
Wako 351-0198, Japan}
\affiliation{$^2 $Center for Mathematics and Physics, University of Aizu, Aizu-Wakamatsu, Fukushsima
965-8560, Japan}

%

\begin{abstract}
We study the sum rules of double Gamow-Teller (DGT) excitations through double spin-isospin operator $(\sigma t_-)^2$.  
In general,  $2^+$ states in the granddaughter nuclei have dominant transition strength in DGT excitations and  $0^+$ states are weak,   except 
 in $T=1$  mother nuclei in which $0^+$ strength is competitive with 2$^+$ strength.  
The isospin selection of DGT is also discussed among 5 possible isospin states in granddaughter nuclei.  
 A possibility to extract  the unit cross section for  the DGT transition strength is pointed out in  the $(\sigma t_-)^2$ excitation of
 double isobaric analog state (DIAS) in $T=1$ nuclei.
\end{abstract}

\maketitle
Double charge exchange excitations (DCX) induced by heavy ion beams at intermediate energies  \cite{RIKEN, INFN-Catania}  
attract a lot of interest in relations with new collective excitations such as double Gamow-Teller giant 
resonance (DGTR) and also double beta decay matrix elements. 
In 1980s, the double charge exchange reactions (DCX) were performed by using pion beams, i.e., 
$(\pi^+, \pi^-)$ and $(\pi^-, \pi^+)$ reactions. 
Through these experimental studies, the double isobaric analog states (DIAS), the dipole giant 
resonance (GDR) built on the IAS and
the double dipole resonance states (DGDR) are identified \cite{Mord-91,Ward-93,Chomaz-95}.  
However, DGTR were not found in the pion 
double charge exchange spectra. 
In the middle of 1990s, the heavy-ion double charge exchange experiments were performed at energies of 
76 and 100MeV/u with a hope that DGTR might be observed in the  $^{24}$Mg($^{18}$O, $^{18}$Ne)$^{24}$Ne reaction \cite{MSU-18O}. 
 However, no clear evidence of DGTR was found in these reactions. 
This is mainly because the  ($^{18}$O, $^{18}$Ne) reaction is a $(2n, 2p)$ type DCX one and even  the single GT $\tau_+$ resonance by $(n,p)$ reaction  is 
weak in a  $N=Z$ nucleus $^{24}$Mg. 
Other  reasons were low intensity of beams and non-optimal energy to excite the spin-isospin responses. 
A new research program based on a new DCX reaction ($^{12}$C, $^{12}$Be(0$^+_2$ )) is planned at RIKEN RIBF facility with high intensity heavy 
ion beams at the optimal energy of E$_{lab}$ =250MeV/u to excite the spin-isospin response \cite{RIKEN-new}. 
 A big advantage of this reaction  is based on the fact that it is 
 a $(2p,2n)$ type DCX reaction and one can use  neutron-rich target to excite DGT strength. 
 Although many theoretical efforts have 
been paid for studies of double beta decays, DGT strengths corresponding 
to the double beta decays are still too small to be identified in the reaction 
experiments. Minimum-biased theoretical prediction based on sum rules 
will  provide a robust and global view of the DGTR and can be a good guideline 
for the future experimental studies.

In this paper, we present useful formulas to analyze the DGT strength using several sum rules for the double spin-isospin operator $(\sigma t_-)^2$. 
We study also DIAS  excited through the double Fermi transition operator $(t_-)^2$  and double GT operator  $(\sigma t_-)^2$.
The DIAS state might be a sharp peak close to DGT peaks and can be used to  extract the unit cross section between the absolute   DGT strength and  the forward angle DCX  cross sections.  The isospin structure of DGT strength  is also given by using the isospin coupling scheme of DGT states.

The sum rule for the single GT transitions  is well known and proportional to the neutron excess,
\bea
&&S_--S_+   \nonumber \\
&=&\sum_f|\bra f|\hat{O}_-({\rm GT})|i\ket|^2 - \sum_f|\bra f|\hat{O}_+({\rm GT})|i\ket|^2 \nonumber \\
&=&3(N-Z)
\eea
where the GT transition operators  read
\be
\hat{O}_{\pm}(\rm{GT})=\sum_{\alpha}\sigma(\alpha) t_{\pm}(\alpha).
\ee
The GT sum rule is model independent and gives a good guidance to perform the single charge exchange 
reactions such as  (p,n) and ($^3$He,t) reactions,  to observe GTR strength in many nuclei (see for example a review article ref. \cite{Sakai-06}). 
  
The DGT transition operator $\hat{O}_{\pm}({\rm GT})^2$ can be projected to good multipole states to be
\be
[\hat{O}_{\pm}({\rm GT})\times\hat{O}_{\pm}({\rm GT})]^{J}_{\mu},  J=0,2
\ee
The sum rule strength is defined as
\bea
D^{J}_{\pm}=\sum_{J_fM_f,\mu}|\bra J_f,M_f|[\hat{O}_{\pm}\times\hat{O}_{\pm}]^{J}_{\mu}|J_iM_i\ket|^2  \nonumber \\
=\bra J_iM_i|\sum_{\mu}(-1)^{\mu} [\hat{O}_{\mp}\times\hat{O}_{\mp}]^{J}_{\mu} 
 [\hat{O}_{\pm}\times\hat{O}_{\pm}]^{J}_{-\mu}  |J_iM_i\ket 
\eea
where $J_i$  and $J_f$  are  the angular momenta of the initial and   final states,respectively.
The sum rule strength is also expressed by the reduced matrix element,  
\be
D^{J}_{\pm}=\frac{1}{2J_i+1}\sum_{J_f}|\bra J_f||[\hat{O}_{\pm}\times\hat{O}_{\pm}]^{J}||J_i\ket|^2,
\ee
where  the double bar means the reduced matrix element for  the angular momentum.  Hereafter we denote the initial state $ |J_iM_i\ket $ by a simple notation $|i\ket$.
The sum rule value for $J=0$ excitations are then evaluated as 
\bea
&&D^{(J=0)}_{-}-D^{(J=0)}_{+}   \nonumber \\
&=&
\bra i|\left[ [\hat{O}_{+}\times\hat{O}_{+}]^{(J=0)}, 
 [\hat{O}_{-}\times\hat{O}_{-}]^{(J=0)} \right] |i\ket  \nonumber \\
&=&2(N-Z)(N-Z+1) \nonumber \\
&+&\frac{4}{3}\left[(N-Z)S_+-\bra i|[i\hat{\Sigma}\cdot(\hat{O}_-\times\hat{O}_+) +\hat{\Sigma}\cdot\hat{\Sigma}]  |i\ket \right] \nonumber \\
\label{eq:D-J-0}
\eea
where
\bea
\hat{\Sigma}=\sum_{\alpha}\sigma(\alpha)   
\eea
and $(\hat{O}_-\times\hat{O}_+)$ is the vector product of two operators.
In the last line of Eq.  (\ref{eq:D-J-0}), we use an identity
\be
(\hat{O}_+\times\hat{O}_-) \cdot\hat{\Sigma}=-\hat{\Sigma}\cdot(\hat{O}_-\times\hat{O}_+) +2i\hat{\Sigma}\cdot\hat{\Sigma}
\label{identity}
\ee
which makes the final formula of the sum rule (\ref{eq:D-J-0})  different from those in the references
\cite{Muto,Vogel}, but it is  equivalent to the one in ref. \cite{Zamick}.  It is important for quantitative  study to rewrite the formula by using 
the identity (\ref{identity}) since the terms including $\hat{O}_{+}$ on the right can be omitted  in $N>Z$ nuclei in a good accuracy.  
The sum rule for $J^{\pi}=2^+ $ final states can be obtained in a similar way to that for $J=0$   states;
\bea
&&D^{(J=2)}_{-}-D^{(J=2)}_{+}   \nonumber \\
&=&
\bra i|\sum_{\mu}(-1)^{\mu}\left[ [\hat{O}_{+}\times\hat{O}_{+}]^{(J=2)}_{\mu}, 
 [\hat{O}_{-}\times\hat{O}_{-}]^{(J=2)}_{-\mu} \right] |i\ket  \nonumber \\
&=&10(N-Z)(N-Z-2) \nonumber \\
&+&\frac{10}{3}\left[2(N-Z)S_++\bra i|[i\hat{\Sigma}\cdot(\hat{O}_-\times\hat{O}_+) +\hat{\Sigma}\cdot\hat{\Sigma}]|i\ket \right] \nonumber \\
\label{eq:D-J-2}
\eea
It should be noticed that the sum rule for  $J=2$ final states are not given in refs.  \cite{Vogel,Zamick}, but given in ref. \cite{Muto} in a different form.
\begin{table*}[hbt]
\caption{Various sum rule values for DGT transitions to $J^{\pi}=0^+, 2^+$ and DIAS states.  The double Fermi transitions to DIAS are also listed.  
For Eqs. (\ref{eq:D-J-0}) and (\ref{eq:D-J-2}), the spin sum rule $\hat{\Sigma}\cdot\hat{\Sigma}$
terms in the last  line is obtained  by shell model calculations: $^6$He, $^8$He and $^{14}$C in  p-shell model space with CKII interaction, $^{18}$O and $^{20}$O in  sd-shell model space with USDB interaction, and Ca isotopes in  pf-shell model space with GXPF1A interaction.  The single particle value of $\Sigma_{sum}$ is used for $^{90}$Zr.  The single particle values from Eq. (13) are given in the bracket in the sixth column.
In the second and third columns, the value in the bracket is 
the upper limit for $J=0$, while  that is the lower limit for $J=2$. Notice that Vogel et al., in ref.  \cite{Vogel} 
and Zheng et al. in ref. \cite{Zamick} 
calculated only $J=0$ channel and a factor 3 larger than the present one because of the exact angular momentum projection in the present results. The values $\Sigma_{sum}$ for DIAS$^{(J=0)}$(DGT) in Eq. (11) are calculated by using the same shell model calculations as for $\Sigma_{sum}$.  
  The sum rule value DIAS$^{(J=0)}$(DGT) with the single particle value $\Sigma_{sum}$ (13)
is also listed in the fifth column with bracket.}
\label{tab1}
\begin{center}
\begin{tabular}{cccccc}\hline
 Initial state    &  $(J=0)$ & $(J=2)$ &  \rm{DIAS}  &   \rm{DIAS(DGT)}  & $\Sigma_{sum}$   \\ \hline
$^6$\rm{He}   & 12.0(12)  & 0.0 (0) & 4   & 3.70(3.70)  &   0.211$\times10^{-3}$ (0.667)   \\
$^8$\rm{He}  &  39.7(40) & 80.7 (80)  & 24 &  7.71(2.47) &   0.052 (1.333)\\
 $ ^{14}$\rm{C}  &  8.98(12)  &  7.55 (0 ) & 4  &4.56 (0.15) & 0.566 (1.333)\\
$^{18}$\rm{O} &  10.4(12) &  3.96 (0)  &4  & 7.72 (2.61)  & 0.297 (0.80)\\
$^{20}$\rm{O} &  35.5(40) & 91.3 (80) & 24  & 4.13 (1.74) &0.845 (1.60)\\
$^{42}$\rm{Ca} & 8.50(12)  &  8.75 (0)  & 4  &     3.80 (2.18)                   & 0.66 (0.86)\\
$^{44}$\rm{Ca} &  32.6(40)  &  98.5(80) & 24  & 2.31 (1.47?j& 1.38 (1.71)\\
$^{46}$\rm{Ca} &   72.3(84)  &  269.3(240)   &   60  &  1.89 (1.32)  & 2.20 (2.57)\\
$^{48}$\rm{Ca} &   135.5(144) & 501.2(480) &112   &    3.70 (1.25)  & 1.59 (3.43) \\
$^{90}$\rm{Zr}&   196.3(220)  & 859.2(800) & 180  &   (1.11) &   (4.44) \\\hline
\end{tabular}
\end{center}
\end{table*}

Next, we consider the DGT transition to the DIAS state.  The DIAS state is  defined as 
\be
|{\rm DIAS}\ket =\frac{T_-T_-}{N_f}|i\ket
\ee
where $T_-=\sum_{\alpha}t_-(\alpha)$  is the isospin lowering operator and the normalization factor is $N_f=\sqrt{2(N-Z)(N-Z-1)}$.
The double Fermi transition to DIAS state is evaluated to be
\be
|\bra {\rm DIAS}|T_-T_-|i \ket|^2=2(N-Z)(N-Z-1) 
\label{DIAS-F}
\ee
in which the mother state is assumed to have a good isospin quantum number $T=T_z=(N-Z)/2$.  
The DGT transition to DIAS is also expressed as 
\bea
&&D^{(J=0)}_-({\rm DIAS})=\left|\bra {\rm DIAS}|\left[\hat{O}_-\times \hat{O}_-\right]^{(J=0)}|i\ket\right|^2 \nonumber \\
&=&\left|\frac{1}{N_f}\bra\hat{0}|T_+T_+\left[\hat{O}_-\times \hat{O}_-\right]^{(J=0)}|i\ket \right|^2   \nonumber \\
&=&\frac{1}{3N_f^2}\left[8\Sigma_{sum}-2(S_-+S_+)\right]^2  \nonumber \\
&=&\frac{1}{6(N-Z)(N-Z-1)}\left[8\Sigma_{sum}-2(S_-+S_+)\right]^2 \nonumber \\
\label{DIAS-GT}
\eea
where $\Sigma_{sum}$ is the sum rule of the isovector spin transition,
\be
\Sigma_{sum}=\bra i|\sum_{\alpha,\mu}\sigma_{\mu}(\alpha)t_z(\alpha)\sum_{\beta, \mu'}\sigma_{\mu'}(\beta)t_z(\beta)|i\ket.
\label{IVspin-sum}
\ee
For a single-j configuration with the occupation probability $v^2$, the value $\Sigma_{sum}$ is evaluated to be
\bea
\Sigma_{sum}&=&\sum_{p,h,\mu}|\bra(ph)1\mu|\sigma_{\mu}t_z|\hat{0}\ket|^2  \nonumber \\
&=&\frac{1}{4}\frac{v_h^2(2j_h+1)(2j_h-1)}{j_h}  \,\,\, {\rm for}  \,\,\, j_h=j_p+1  \nonumber \\
\label{spin-sum-sp}
\eea
where we use a filling approximation for the last occupied orbit.  

Calculated sum rule values for DGT transitions to $J^{\pi}=0^+$ and  $2^+$
in Eqs. (\ref{eq:D-J-0}) and (\ref{eq:D-J-2}), respectively,  are listed in Table 1 for 10 nuclei, together with those for  DIAS states by Fermi and GT operators (\ref{DIAS-F}) and (\ref{DIAS-GT}).   In $N>Z$ nuclei with the good isospin quantum number, the vector product $(\hat{O}_-\times\hat{O}_+)$ and the sum $S_+$ in Eqs. (\ref{eq:D-J-0}) and (\ref{eq:D-J-2}) vanish because of the isospin raising operator in 
$\hat{O}_+$   
.   The spin sum rule $\hat{\Sigma}\cdot\hat{\Sigma}$
term is obtained  by shell model calculations: $^6$He, $^8$He and $^{14}$C with p-shell model space with CKII interaction, $^{18}$O and $^{20}$O with sd-shell model space with USDB interaction, and Ca isotopes with pf-shell model space with GXPF1A interaction.  The single particle value is used for $^{90}$Zr assuming the neutron excitation from
 $1g_{9/2}$ to $1g_{7/2}$.  
The value in the bracket is obtained without the spin sum rule term, i.e., 
the upper limit for $J=0$, while  the lower limit for $J=2$. Notice that Vogel et al.,  in ref. \cite{Vogel} 
and Zheng et al.  in ref. \cite{Zamick} 
calculated only $J=0$ channel and give a factor 3 larger than the present one because of no  exact angular momentum projection in their results. The values $\Sigma_{sum}$ 
for DIAS$^{(J=0)}$(DGT) in Eq. (\ref{DIAS-GT}) are calculated by using the same shell model  wave functions as for  $\Sigma\cdot\Sigma$ 
and shown in the  sixth column in Table 1.  The single particle values of Eq. (13) are given in the bracket in the sixth column. The value DIAS$_-^{(J=0)}$(DGT) with the single particle value $\Sigma_{sum}$
is  listed in the fifth column with bracket. In all the nuclei in the present study, the value 
of $\Sigma_{sum}$ is one-fourth of $\hat{\Sigma}\cdot\hat{\Sigma}$ because of semi-closed configuration (either proton or neutron configuration is a closed shell).  

The values $\Sigma_{sum}$, equivalently  $\hat{\Sigma}\cdot\hat{\Sigma}$, are almost negligible for p-shell nuclei $^{6}$He and  $^{8}$He, 
because of good SU(4) spin-isospin symmetry in which the spin transitions are allowed only for spin and  isospin flip transition between $(S, T)=(0,1)\rightarrow(1,0)$.  This selection turns out  
to be  the selection  of  transitions $(J^{\pi}, T)=(0^+,1)\rightarrow(1^+,0)$ in $^{6}$He and $(J^{\pi}, T)
=(0^+,2)\rightarrow(1^+,1)$  in  $^{8}$He,  which correspond to the charge exchange GT transitions.  
The shell model values $\Sigma_{sum}$ are smaller than the single-particle values in other nuclei by a factor $1.2\sim 2.5$.  The effect of $\hat{\Sigma}\cdot\hat{\Sigma}$ term for $J=0$  and $J=2$ sum rules are negligible for $^{6}$He and  $^{8}$He, while it amounts at most 20\% 
in other nuclei with $T>1$ and it becomes much larger in  $T=1$ nuclei such as $^{14}$C, $^{18}$O and $^{42}$Ca.  

In $N>>Z$ nuclei, the sum rule values of DGT transitions are approximately  proportional to 
(2J+1) factor, i.e., the value for $J=2$ is 5 times larger than that for $J=0$ in the same nucleus.  However, this proportionality is strongly modified in $N\sim Z$ nuclei.  In the extreme,  the sum rule for $J=2$ transitions is smaller than that for $J=0$ in the nuclei $^6$He, $^{14}$C, and $^{18}$O with$ N=Z+2$ , and  the two values are almost equal in $^{42}$Ca.  

It is noticed that the DGT transition to DIAS state is much smaller than the sum rule values of DGT transitions for 
 $J=0$ and $J=2$ states  in nuclei with $N>Z+2$.  However, in $T=1$ nuclei, the DGT transition to DIAS state is comparable with the DGT sum rules (\ref{eq:D-J-0})  
  and  (\ref{eq:D-J-2}).  This characteristics 
turns out the possibility to extract the unit cross section for DGT transition in $T=1$ nuclei,
 especially $^{14}$C and $^{18}$O,    
since the DIAS state might have a narrow width and may be distinguished clearly from other DGT states 
in double charge exchange reactions.   This argument is entirely benefitted from the fact that  the spin-isospin response is much favorable that the isospin response in the medium energy charge exchange reactions with $E_{lab}\sim $200MeV/u.

So far the isospin dependence of DGT sum rules in  granddaughter nucleus is not considered.
For the DGT transitions from a  mother nucleus with the isospin $T$,  five different isospin states $T''=T+2, T+1, T, T-1, T-2$ are expected for the granddaughter nucleus  with $T_z=T-2$.
DIAS is a state with $T''=T$.  The isospin selection amplitude $A(T'')$ for different isospin $T''$ in granddaughter nucleus can be calculated by a combination of Clebsch-Gordan coefficients,
\bea
&&A(T^{''})  \nonumber \\
&=&\sum_{T'}\bra TT1-1|T'T-1\ket \bra T'T-11-1|T''T-2\ket  \nonumber \\
       &=&\bra TT2-2|T''T-2\ket  \nonumber \\
      &=&   \left\{
    \begin{array}{cc}
      \sqrt{\frac{2T-3}{2T+1}} &  for \,\,\,\, T{''}=T-2 \\
      \sqrt{\frac{2(2T-1)}{(T+1)(2T+1)} }    & for  \,\,\,\, T{''}=T-1 \\
     \sqrt{\frac{6}{(T+1)(2T+3)}  }   & for  \,\,\,\, T{''}=T \\
     \sqrt{\frac{6}{(T+1)(T+2)(2T+1)}}     & for   \,\,\,\,  T{''}=T+1 \\
      \sqrt{\frac{6}{(T+1)(T+2)(2T+1)(2T+3)}  }   & for  \,\,\,\, T{''}=T+2 
   \end{array}
  \right\}  \nonumber \\
\eea
The calculated values referred to  mother nuclei with different isospins $T=1\sim 22$ are given in Table II.
For the mother nucleus with $T=1$, only three isospin states $T''$ are allowed by DGT excitations.  For mother nuclei with large isospin $T>2$, the isospin $T''=T-2$ give the major contributions.  On the other hand,  for mother nuclei with small isospin $T\leq$2, the dominant configurations are $T''=T$ for $T=1$ and $T''=T-1$ for $T=2$ mother nuclei.

\begin{table}[t]
\caption{Splitting of DGT strength to different isospin final state with $T''=T\pm2,T\pm1,T$ for the initial state with $T=1\sim 22$. Probability  $A(T'')^2$  of  each isospin $T''$ for a given $T$ is  listed  in \% 
.}
\label{tab2}
\begin{center}
\begin{tabular}{c|ccccc}\hline
 $T$  & \multicolumn{5}{c}{$T''$}  \\ \hline
 &  $T-2$ &  $T-1$ & $T$ & $T+1$  &$T+2$  \\ \hline
1 &   -    &  - &   60  &  33.3  &   6.7  \\
2  &  20  &  40  &  28.6  &  10  &  1.4   \\
3  &  42.9  &  35.7  &   16.7  &   4.3  &  0.5  \\
4  &   55.6  &  31.1  &  10.9  &  2.2  &  0.2  \\
 5  &  63.6  &  27.3  &  7.7  &  1.3  &  0.1  \\
  8  &  76.5  &  19.6  &  3.5  &   0.4  &   0.0  \\
  22 &  91.1  &  8.3  &  0.6  &   0.0  &   0.0   \\
\hline
\end{tabular}
\end{center}
\end{table}

In summary, we studied the sum rules of DGT excitations by the operator $(\sigma t_-)^2$.   In nuclei $N>Z+2$, the $J^{\pi}=2^+$ excitations dominate the DGT strength,  due to 
the multipole factor $(2J+1)$,  more than $0^+$ excitations.  However, in nuclei with $N\sim Z$, the $0^+$ excitations become substantially strong,  even larger than 
2$^+$ excitations in light nuclei with  $T=1$ such as $^{14}$C an $^{18}$O.  The excitation to DIAS is also studied through $(\sigma t_-)^2$ and $t_-^2$ operators 
to investigate the possibility to extract the unit cross sections for DGT strength.  We pointed out that the strength of DIAS excitations by $(\sigma t_-)^2$ operator 
is competitive to DGT strength in the light $T=1$ nuclei.  This characteristic feature  may give a good opportunity to extract the unit cross section for DGT strength.  
The isospin structure of DGT is also studied in the granddaughter nuclei.  Among 5 possible isospin states $T''=T\pm2, T''=T\pm1, T''=T$ in  a granddaughter nucleus referred to a  mother state 
with the isospin $T$,  the largest component appears for $T''=T-2$ states for $T>2$ nuclei,  while $T''=T$ and $T''=T-1$ states are dominant in the final states 
in $T=1$ and $T=2$ nuclei, respectively.  

\section*{Acknowledgments}

This work was supported in part by JSPS KAKENHI  Grant Numbers JP16K05367.
{}
\end{document}